\documentclass[9pt,twocolumn,twoside]{opticajnl}

\journal{opticajournal} 

\setboolean{shortarticle}{false}

\usepackage{lineno}

\usepackage{braket}

\title{Efficient detection of multidimensional single-photon time-bin superpositions}

\author[1,2,*]{Adam Widomski}
\author[1,2]{Maciej Ogrodnik}
\author[1]{Micha{\l} Karpi\'nski}
\affil[1]{Faculty of Physics, University of Warsaw, Pasteura 5, 02-093 Warszawa, Poland}
\affil[2]{The authors contributed equally to this work.}
\affil[*]{at.widomski@uw.edu.pl}

\begin{abstract}
The ability to detect quantum superpositions lies at the heart of fundamental and applied aspects of quantum mechanics. The time-frequency degree of freedom of light enables encoding and transmitting quantum information in a multi-dimensional fashion compatible with fiber and integrated platforms. However, the ability to efficiently detect time-frequency superpositions is not yet available. Here we show, that multidimensional time-bin superpositions can be detected using a single time-resolved photon detector. Our approach uses off-the shelf components and is based on the temporal Talbot effect -- a time-frequency counterpart of the well-know near field diffraction effect. We provide experimental results and discuss the possible applications in quantum communication, quantum information processing, and time-frequency quantum state tomography.
\end{abstract}

\setboolean{displaycopyright}{false}

\begin{document}

\maketitle

\section{Introduction}

The ability to detect quantum superpositions is key for most of quantum experiments, from quantum communication \cite{BB84, E92, Bechmann2000, Cerf2002}, quantum state and process tomography \cite{ParisRehacek2004, ansari2018tomography, Lu2022, Kues2017, serino2023, Jack_2009, riebe2006, steffen2006}, entanglement verification \cite{bavaresco2018measurements, guo2018witness, Schneeloch2018EPR, guhne20091, hyllus2005relations} to quantum computation \cite{nielsen2010quantum}.
Given the growing use of the time-frequency degree of freedom of quantum light, which enables multidimensional encoding of quantum information in a single spatial mode, compatible with fiber optic and integrated optical setups \cite{Brecht2015, karpinski2021control, Lukens2017, Roztocki:19, roslund2014wavelength}, a necessity arises to efficiently detect single-photon superpositions in time and frequency \cite{bruss2002optimal, sheridan2010, karpinski2021control, Brecht2015}.

The measurement of temporal mode superpositions was performed by means of the optically nonlinear quantum pulse gate \cite{Eckstein2011, Reddy2018}, enabling quantum tomography \cite{Ansari2017}, and by means of electro-optic sideband generation in frequency bins, enabling time-frequency entanglement detection \cite{Kues2017, Lu2022, Lukens2017}.
However, those approaches require active spectral modification, either electro-optically or by nonlinear optical interactions.
Franson interferometers \cite{franson1989bell, Zhong2015} are an alternative for the detection of time-bin superpositions, but a single interferometer detects the phase between only two time bins.
Building a setup enabling multidimensional state discrimination requires nesting the interferometers \cite{islam2017robust, brougham2013security}.
This implies higher complexity, cost, and in particular, losses. In the Franson interferometer tree method the probability of detecting a $d$-dimensional superposition scales as $1/d$, due to the increasing number of possible paths in the nested interferometer \cite{islam2017robust}. 

Another possibility to approximately detect time-bin superpositions is by directly resolving their spectrum. However, resolving the signals in time and frequency simultaneously at the single-photon level is challenging due to the limits given by the time-frequency uncertainty relation \cite{karpinski2021control, widomski}. The highest resolution of efficient multiplexed single-photon spectral measurements is offered by means of the dispersive Fourier Transform \cite{goda2013dispersive, Davis2017, Avenhaus2009, Narhi2018, kolenderski2020}, however, the large values of group dispersion delay needed to reach the required spectral resolution can result in prohibitively high losses \cite{karpinski2021control}.

Here we show, that the temporal Talbot effect \cite{jannson1981temporal}, a time-frequency equivalent of the well-known near field diffraction effect \cite{talbot1836}, enables discrimination between $d$ orthogonal states of a $d$-dimensional time-bin superposition. Our method only requires a passive experimental setup comprising a dispersive medium and a single time-correlated single photon counter. We show, that whereas the discrimination is not perfect, the information content increases with the dimension. The setup can be constructed with off-the-shelf components, requires significantly less dispersion than a conventional dispersive Fourier spectrometer, and offers constant detection loss regardless of the dimension $d$.\\

\section{Methods}
\subsection{Temporal Talbot effect}
\begin{figure}[h!]
\includegraphics[scale=1.0, width=\columnwidth]{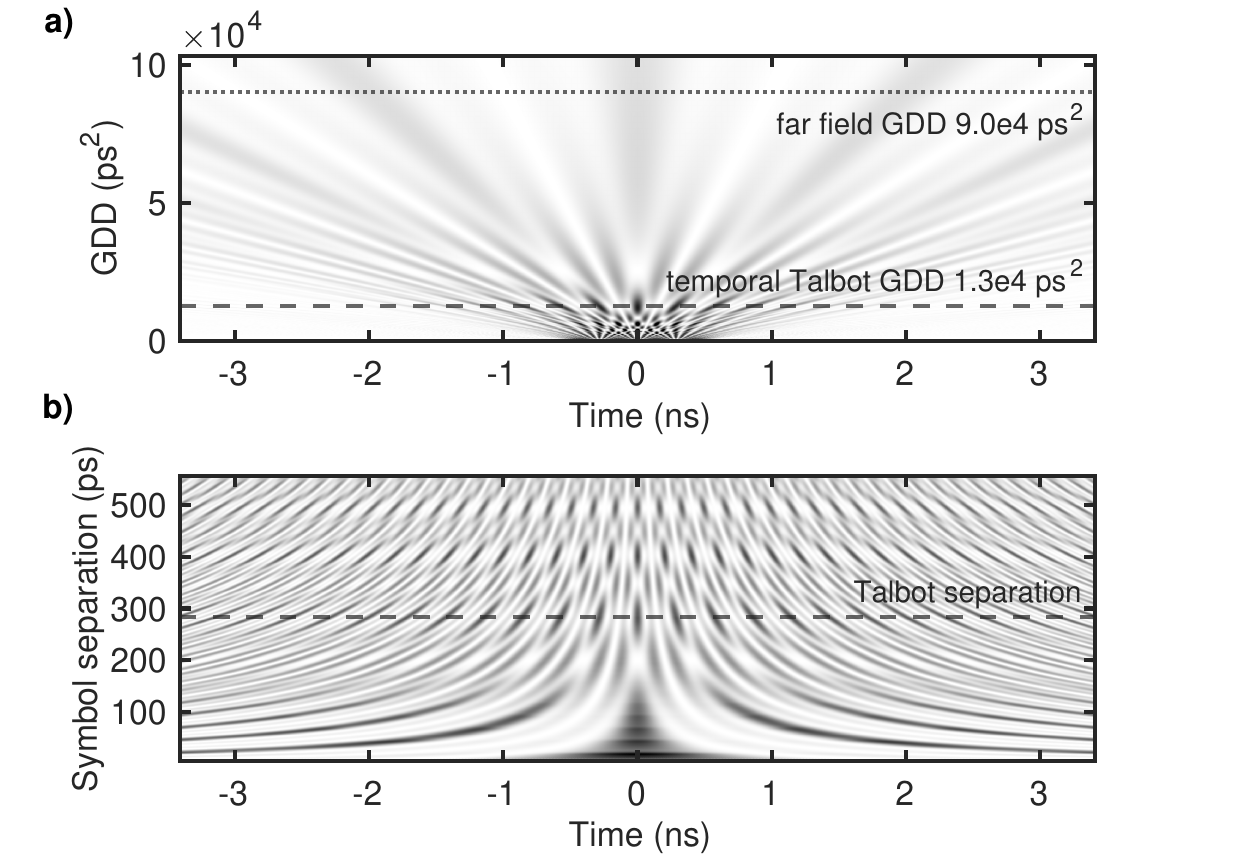}
\caption{a) A Talbot carpet displaying probability of finding a photon at a specified instance of time as a function of group delay dispersion (GDD). Color intensity depicts the probability density. Horizontal dashed lines represent GDD values necessary to measure the distribution in the temporal far field and by means of the temporal Talbot effect. The carpet was generated considering no phase difference between the pulses.
b) An orthogonal Talbot carpet representing the probability to detect a photon in time as a function of symbol separation, calculated for $\textrm{GDD} = 12900 \textrm{ ps} ^2$ and  $\tau=284$ ps corresponding to the first Talbot separation. 
\label{fig:simulation}}
\end{figure}

The Talbot effect is a phenomenon of self-imaging occurring for periodically placed intensity distributions \cite{talbot1836}.
Those self-images are created due to diffraction, and repeat themselves after a certain characteristic distance.
The optical space-time duality \cite{kolner1994space} lets us translate spatial effects within the paraxial approximation to temporal effects in short pulse propagation in a second-order dispersive medium, in the narrowband regime \cite{Torres2011}.
This means that a superposition of equally spaced pulses is self-imaged by propagation in a medium with a certain group dispersion delay (GDD) (which we will further refer to as the Talbot GDD) \cite{jannson1981temporal}, just as equally spaced pulses would be self-imaged after some propagation distance due to diffraction (Fig. \ref{fig:simulation}a). This temporal Talbot effect enables multiple bright light applications such as multiplication of the repetition rate of a periodic pulse train \cite{azana2001temporal}, factorization of numbers \cite{bigourd2008factorization} or real-time calculation of microwave spectrograms \cite{azana2023optical}.

In analogy to the spatial case, the temporal location of the self-images depends on the relative phases of the sources. This enables us to employ the temporal Talbot effect to detect different time bin superpositions. In particular, we use the fact that the temporal Talbot effect can be seen as a discrete Fourier transform of input pulse trains \cite{tainta2014temporal, fernandez2017dispersion, zhou2010all}. 

\subsection{Detecting superpositions}
Let us consider optical pulses in 4 time bins:
\begin{equation}
    \ket{t_0}, \ket{t_1}, \ket{t_2}, \ket{t_3}.
\end{equation}

Those time bins are approximately Gaussian-shaped, $40 \textrm{ ps}$ wide and separated by $284 \textrm{ ps}$ (the experimental details will be presented in sec. 2c.).

Those states can be directly measured in time by means of time-correlated single photon counting (TCSPC). We also created 4 superpositions of those states that form a discrete Fourier transform \cite{barnett2009quantum} of the time bin basis:
\begin{equation}
    \ket{f_n} = \frac{1}{\sqrt{d}} \sum_{m=0}^{d-1} e^{\frac{-2\pi i n m}{d}} \ket{t_m},
    \label{eq:dft-basis}
\end{equation}
with the dimension $d=4$.

Let us now consider propagation through a medium with group delay dispersion corresponding to the Talbot GDD ($\beta_2 $). The Talbot GDD is linked to the separation between the pulses ($\tau$) \cite{azana2001temporal} as:
\begin{equation} \label{eq:TalbotGDD}
    \tau=\sqrt{\frac{2\pi\beta_2}{s}}, s = 1,2... .
\end{equation}
Finally, the signals are detected by time-resolved single photon counting at the output of the GDD medium. When the condition in Eq. \ref{eq:TalbotGDD} is met, a train of pulses in time will be observed (cf. the dashed line in Fig. \ref{fig:simulation}a due to the temporal Talbot effect.

In Fig. \ref{fig:simulation}b we show the dependence of the output time-resolved photon detection probabilities on the symbol separation for a given GDD value of $12900$~ps$^2$, corresponding to the first Talbot separation in the experiment.
Fringe pattern can be observed for symbol separations matching the conditions.

\begin{figure}[h!]
\includegraphics[scale=1.0, width=\columnwidth]{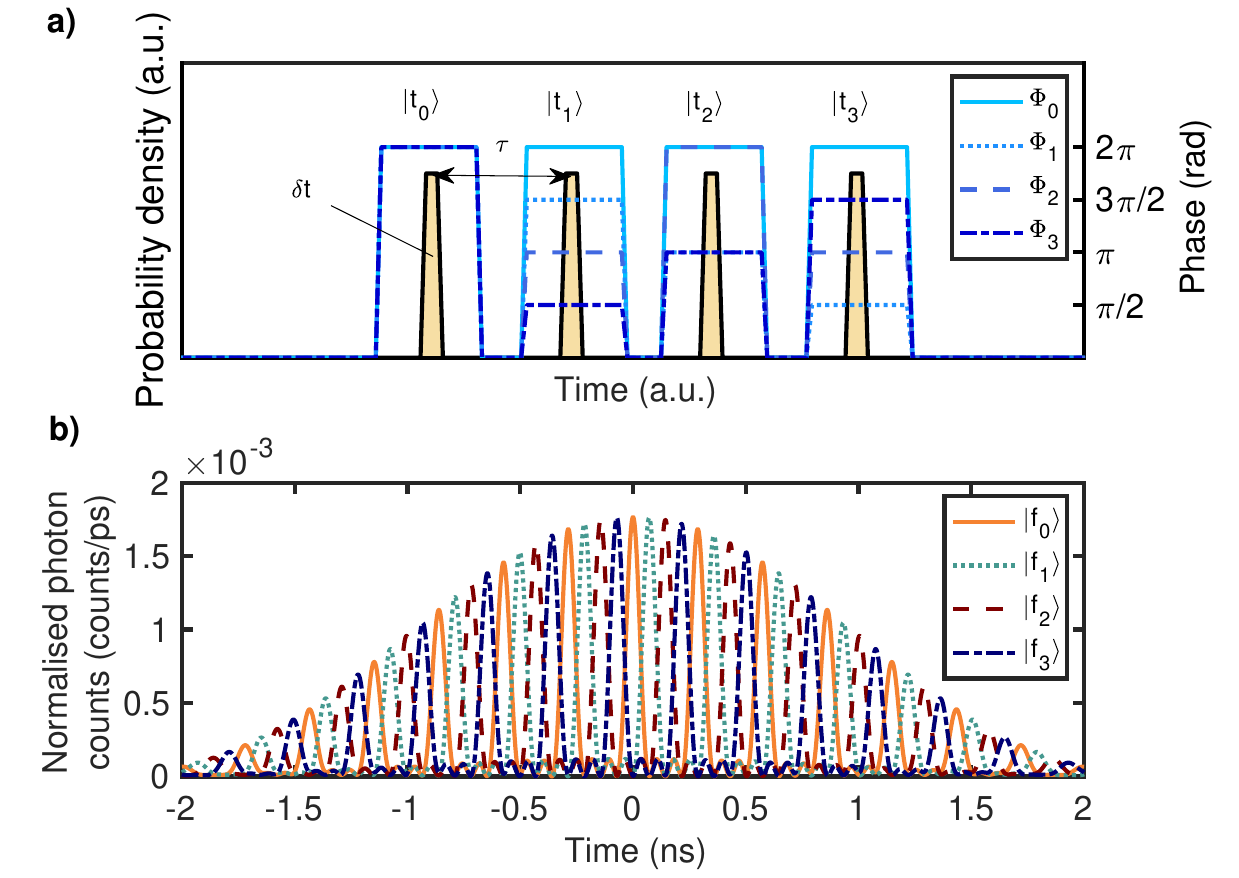}
\caption{a) Numerical calculation of signal's temporal profiles. Yellow rectangles represent temporal symbols generated by means of fast electro-optic amplitude modulation. Blue rectangles represent voltage signals used for phase modulation. b) Simulated single-photon time of arrival distributions for the temporal Talbot effect for the experimental parameters provided in the text. \label{fig:signals}}
\end{figure}

In Fig. \ref{fig:signals}a we present the optical pulses representing the 4 time bins. Those pulses were further used to form superpositions. Adequate phases are adjusted with the voltage signals (blue rectangles) driving an electro-optic phase modulator. The histograms corresponding to the photon times of arrival of the states from the basis $\{\ket{f_i} \}$ are presented in the Fig. \ref{fig:signals}b.

In our method only a single detector is used, and all detections are taken into account, but for a fraction of detections the result may be ambiguous. This is in contrast to the Franson interferometer tree method \cite{brougham2013security} where for a fraction of detections we get unambiguous measurement. 
For single-shot state discrimination in Talbot effect method we need a procedure to assign symbol to a measured time of arrival.
We want to do it with maximal \emph{correctness}, that is a ratio of correct outcomes to errors.
We could try to do that by some probabilistic procedure based on conditional probabilities of each symbol for each time of arrival. Next we show that the optimal strategy is to simply take the most probable symbol at a given time.

With a fixed experimental setup this is a task of distinguishing classical random variables which is a simple case of Bayesian decision theory \cite{duda2006pattern}.
Let us denote the random variable associated with the symbol $i$ as $X_i$.
Symbol number $i$ is chosen randomly, described by the random variable $\sigma$ with the uniform distribution on the $\{0, 1, \ldots , d-1 \}$.
Next, a random time of arrival value $X_i=t_0$ is drawn.
Our task is to decide from which $X_i$ a value $t_0$ was drawn.
Here we consider case with discrete time which corresponds to the realistic setup with a finite time resolution.
We want to guess the outcome of $\sigma$ given the outcome $t_0$ of $X_{\sigma}$.
Based on probabilities $P(X_i = t_0)$ we may decide the outcome probabilistically.
Given the result $t_0$ we give answer $i$ with a probability $\alpha_{t_0} (i)$.
We want to maximize the correctness, ratio of correct results to wrong results, given by
\begin{equation}
    C = \sum_t \sum_i \alpha_t (i) P(\sigma = i \vert X_{\sigma} = t).
\end{equation}
By the Bayes rule we have
\begin{equation}
    P(\sigma = i \vert X_{\sigma} = t) = \frac{P(X_{\sigma} = t \vert \sigma = i) P(\sigma = i)}{P(X_{\sigma} = t)} = \frac{P(X_{i} = t) \frac{1}{d}}{P(X_{\sigma} = t)},
\end{equation}
so the function to optimize takes the form
\begin{equation}
    C = \sum_t \frac{1}{d \cdot P(X_{\sigma} = t)} \sum_i \alpha_t(i) P(X_i = t).
\end{equation}
We have
\begin{equation}
    \sum_i \alpha_t(i) P(X_i = t) \leq \max_j P(X_j = t) \sum_i \alpha_t(i) = \max_j P(X_j = t),
\end{equation}
and this bound is saturated for the deterministic strategy where $\alpha_t (i) = 1$ for $i$ such that $P(X_i = t)$ is maximal.
For this game the optimal strategy is to always choose the outcome $i$ such that
\begin{equation}
    P(X_i = t_0) = \max_j P(X_j = t_0) .
    \label{corr:criterion}
\end{equation}
Depending on the application it may be advantageous to have lower error rate at the cost of lower efficiency. We can do that in the postprocessing by taking only measurement times at which one symbol is much more probable than others. 

\subsection{Experimental setup}
\begin{figure}[h!]
\includegraphics[scale=1.0, width=\columnwidth]{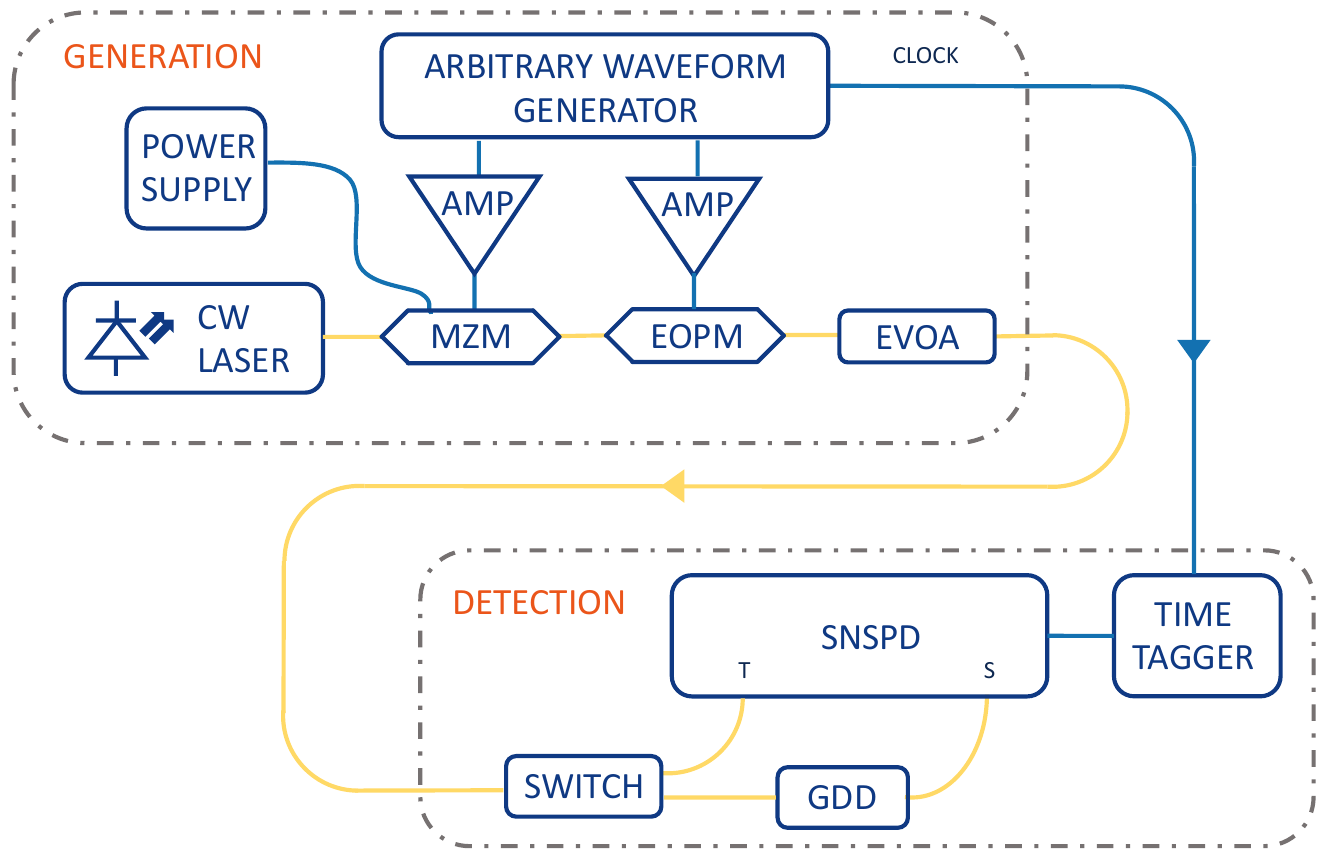}
\caption{Experimental setup. A continuous wave (CW) telecommunications laser is modulated with an electro-optic Mach-Zehnder modulator (MZM) and electro-optic phase modulator (EOPM) to generate optical pulses forming superpositions. The optical signals are attenuated to the single photon level with an electronic variable optical attenuator (EVOA). The signals are detected either directly in the time domain (T) or in the superposition domain (S) by means of the temporal Talbot effect performed with the dispersion compensating module (DCM) providing group delay dispersion (GDD). Photon time-of-arrival histograms were acquired by means of time-correlated single photon counting (TCSPC). Yellow lines represent optical fiber connections. \label{fig:setup}}
\end{figure}

Our experimental setup is shown in Fig. \ref{fig:setup}.
A laser operating at the telecom wavelength of 1560 nm in the continuous wave (CW) mode was used as source of light.
For technical reasons, we amplified the optical signal with an erbium-doped fiber amplifier (EDFA, Pritel, HPP-PMFA-22-10) and employed a bandpass filter to reduce the noise. Optical pulses were carved by means of fast amplitude electro-optic modulation with a Mach-Zehnder modulator (MZM, Thorlabs, LNA6213). Those pulses were used to form  superpositions consisting of $d=4$ single states in a predefined phase relation.The superpositions were directed to the electro-optic phase modulator (EOPM, EOspace), which was used to impose adequate phases on every component of the superposition.
Both modulators support 40 GHz of analog bandwidth, and were driven with an amplified radio-frequency (RF) signal generated with the fast arbitrary waveform generator (AWG, Keysight, M1896A) providing 33 GHz of the analog bandiwdth and sampling rate up to 92.16 Gs/s.
The full-width at half maximum (FWHM) duration of a one-bit signal was therefore $\sim\! 12  \textrm{ ps}$. The phase factors were adjusted with the EOPM by programming four driving voltage signals consisting of approximately $150$-ps-wide rectangular pulses (Fig. \ref{fig:signals}a) such, that their amplitudes corresponded to fractions or multiples of the half-wave voltage (${V_\pi}$). To generate optical signals we assigned 3 bits from the AWG memory per symbol. The separation between the symbols was equal to 282 ps insted of 284 ps due to the finite sampling frequency. The AWG was also a source of the 10 MHz clock (CLK) signal distributed over an electrical cable to the detection section.
We biased the MZM for minimal transmission with the direct current (DC) voltage from the power supply (Keysight, E36313A) using the feedback loop consisting of a 90:10 fiber optic beam splitter and a power meter (Thorlabs, PM400).
Finally, the pulses were attenuated to the single photon level with the electronic variable optical attenuator (EVOA, Thorlabs, EVOA1550F).

On the detection side, we reconnected the optical path in order to either measure the photon counts directly in time, or to detect the superpositions. Time-domain measurement was achieved by directing the signals to niobium nitride superconducting nanowire single-photon detectors (SNSPDs, Single Quantum) and histogramming the time tags using a time-to-digital converter (Swabian Instruments, Time Tagger Ultra).
The time tagger exhibited jitter equal to $\sim\! 10 \textrm{ ps}$ root mean square (RMS), and the SNSPDs of $\sim\! 5\textrm{ ps}$ RMS.
Thus, the total jitter of the receiver system was $\sim\! 11 \textrm{ ps}$ RMS.
For the superposition measurement, photons were additionally transmitted through a chirped-fiber-Bragg-grating-based dispersion compensating module (DCM), resulting in dispersive stretching of the pulses.
We used a DCM providing group delay dispersion (GDD) equivalent to $12900 \textrm{ ps} ^2$ (Proximion, DCMHDC-100H-P510). This amount of dispersion was not enough to perform the dispersive Fourier transform, but sufficed to observe the temporal Talbot effect (see Fig.\ref{fig:simulation}a), which we used to distinguish superpositions of the optical signals with different phases.

\section{Results}

\begin{figure}[h!]
\includegraphics[scale=1.0, width=\columnwidth]{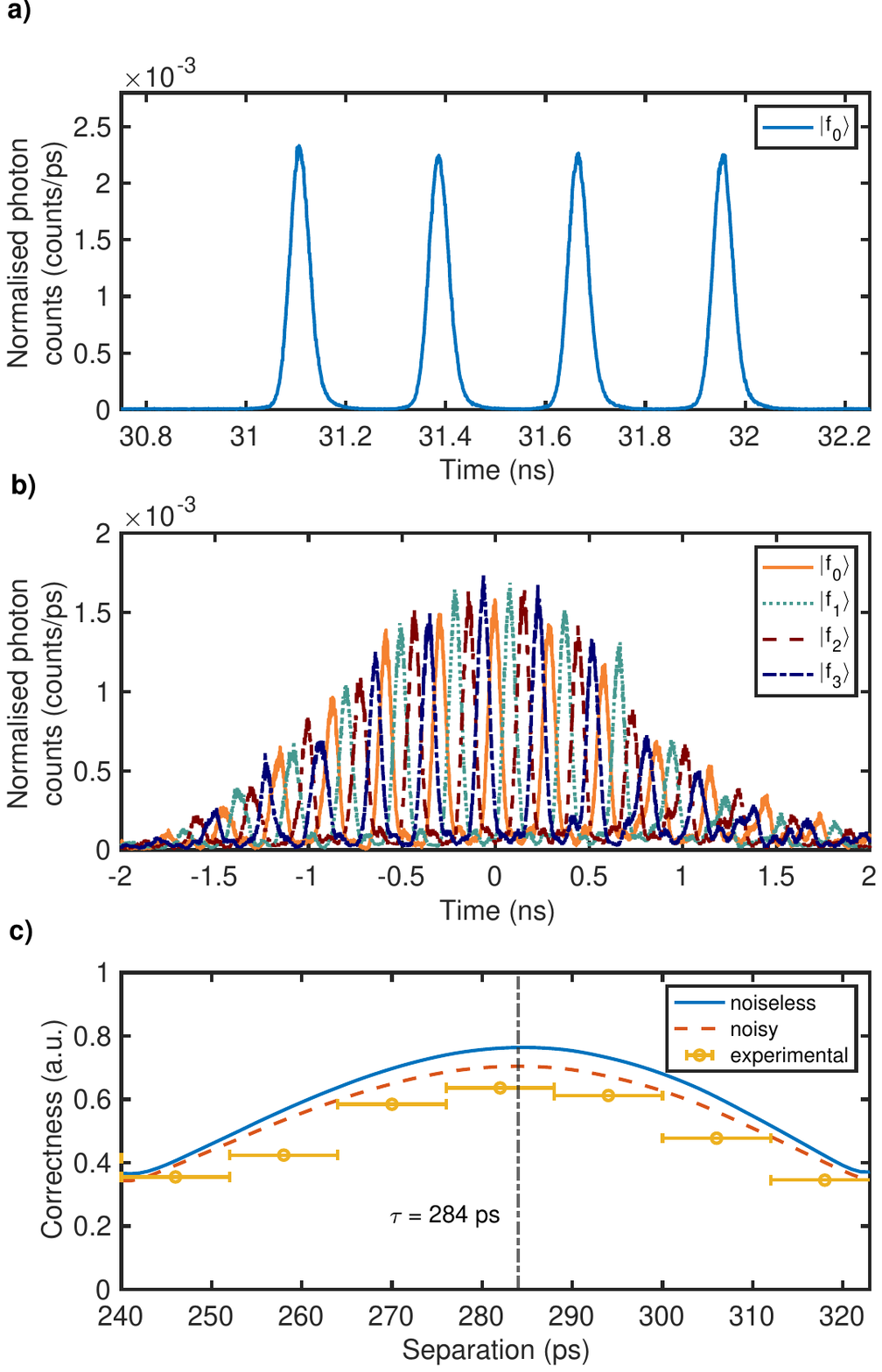}
\caption{a) Measured histogram of the $\ket{f_0}$ state in the time domain. b) Measured histograms of the $\ket{f_i}$  states after frequency-to-time mapping c) Measured correctness of the $\ket{f_i}$ state discrimination, compared to the theoretical dependencies taking into account no jitter (noiseless), and 11 ps root mean square (RMS) jitter (noisy) cases. The error bars indicate sampling error due to the finite sampling frequency of the AWG. Imperfections in symbol preparation (not included in the simulation) lead to a small decrease in the measured correctness.
\label{fig:results}}
\end{figure}

We demonstrated our technique with four four-dimensional superpositions. We analysed histogram's widths, shapes, and locations on the time axis to verify our method. Temporal symbols were approximately-Gaussian, 46 ps wide, and separated by approximately 284 ps, as expected. We examined these temporal properties by utilising the $\ket{f_0}$ state, since all the components of this superposition were equally modulated in phase (Fig \ref{fig:results}a). Fig \ref{fig:results}b presents histograms of all the $\{f_i \}$ states after propagation through the DCM. The obtained fringes were distinguishable in time yielding the possibility to detect each state from the $\{f_i \}$ basis. 

To display the essence of utilising the Talbot effect for efficient detection, we measured the probability of successful measurement for symbol separations lower and greater than the first Talbot separation. The outcomes of this measurement are presented in Fig. \ref{fig:results}c.
We evaluated the correctness, according to the aforementioned selection criterion, eq. \ref{corr:criterion}.
The result indicates that any deviation from the Talbot separation results in the decrease of the correctness. This indicates that the measurements are performed outside the temporal far field regime. 

\section{Conclusion and discussion}
We propose a new method to detect time-bin states from discrete Fourier transform basis. Those states could be also detected by a tree of Franson interferometers as proposed in \cite{brougham2013security, Islam2017}.
For a system of dimension $d$ this method requires $d-1$ aligned Franson interferometers and $d$ detectors and uses postselection of results that discards $\frac{d-1}{d}$ of measurements.
Real efficiency is further limited by the insertion loss of interferometers in the optical path.
In our method only one dispersion module and one detector are used for the measurement.
It offers constant efficiency irrespective of the dimension, at the cost of error rate increasing with the number of different states to be distinguished.
One can recognize a trade-off between the Franson interferometer tree method with a constant error rate, decreasing efficiency and increasing complexity, and our method with a constant efficiency, constant complexity and error rate increasing with dimension.
The error rate in our method can also be decreased by postselection of the results without the change in the experimental setup.

\begin{figure}[h!]
\includegraphics[scale=1.0, width=\columnwidth]{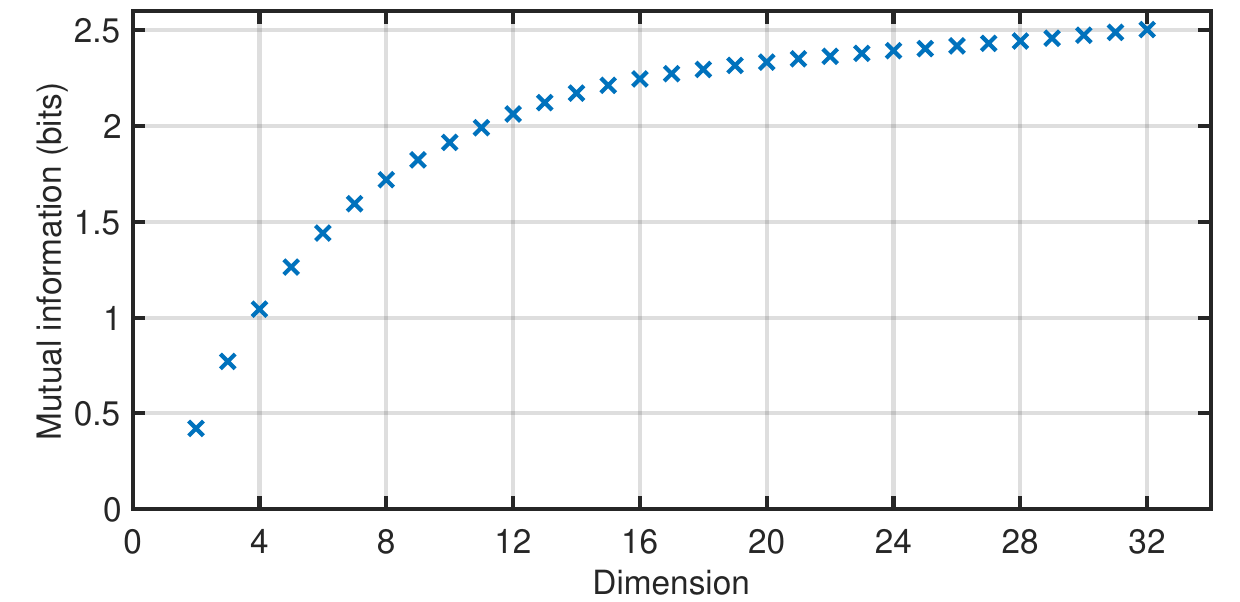}
\caption{Numerical simulation of the mutual information of input state and (noisy) time of arrival random variables.}
\label{fig:mutual_information}
\end{figure}

With larger alphabet we send more bits of information, but we also witness higher error rates.
To quantify that trade-off we may use mutual information to evaluate the scaling with dimension of the amount of information that can be sent using the Talbot effect method for detection.
In particular, we calculate mutual information between the random variable $\sigma$ of the prepared state and the random variable of the time of arrival $X_{\sigma}$.
In Fig. \ref{fig:mutual_information} we can see how it increases with the increase of the dimension of the prepared superposition.

In summary, we show that the temporal Talbot effect in the single-photon-counting regime enables efficient detection of time-bin superpositions in an all-fiber setup.
The proposed selection algorithm allowed us to distinguish the superpositions for all detected photons with with error rate of $36.5 \%$ for the Talbot separation between pulses.
Compared to the Franson interferometer tree method we have higher efficiency at the cost of higher error rate.
Other post selection strategies, including rejection of ambiguous measurements, can alter that trade-off. 
Our technique can be easily expanded to higher dimensions with the increase in bits of information per pulse.
It offers low complexity and high flexibility, which is key for quantum-photonic applications.
We expect it will enable development of robust and efficient quantum communication methods, as well as techniques for time-frequency entanglement detection and characterization and quantum state and process tomography.

\begin{backmatter}
\bmsection{Funding}
A part of this work was carried out within the Project QuICHE, supported by the National Science Centre of Poland project no. 2019/32/Z/ST2/00018, under QuantERA, which has received funding from the European Union's Horizon 2020 research and innovation programme under grant agreement no 731473. The authors appreciate funding from the University of Warsaw within the of the "Excellence Initiative – Research University" framework.
\bmsection{Acknowledgments}
We thank M. Miko{\l}ajczyk, R. Demkowicz-Dobrza\'nski and F. So\'snicki for insightful discussions.

\bmsection{Disclosures}
The authors declare no conflicts of interest.

\end{backmatter}

\bibliography{sample}

\bibliographyfullrefs{sample}

\end{document}